\documentclass[12pt,preprint]{aastex}
\begin{document}
\title{Intrinsic Inclination of Galaxies embedded in Cosmic Sheets 
and Its Cosmological Implications: An Analytic Calculation}
\author{\sc Jounghun Lee}
\affil{School of Physics, Korea Institute for Advanced Study, 
Seoul 130-722, Korea}
\email{jounghun@kias.re.kr}
\received{2004 August 13}
\accepted{2004 ???}

\begin{abstract} 

We investigate analytically a large-scale coherence in the orientation of 
galaxies embedded in two-dimensional sheet-like structures in the frame of 
the tidal torque theory. Assuming that the galaxy spin and the surrounding 
matter fields are intrinsically aligned in accordance with the tidal torque 
model, we first derive analytically the probability distribution of the 
galaxy position angles, and evaluate the degree of their inclinations 
relative to the plane of the sheet.  Then, we apply our analytic approach 
to the nearby spirals in the Local Super Cluster, and provide theoretical 
explanations about why and to what degree the nearby spirals are inclined 
relative to the supergalatic plane. Finally, we conclude that the observed 
large-scale coherence in the orientation of nearby spirals relative to the 
supergalactic plane can be quantitatively understood in terms of galaxy 
intrinsic alignment predicted by the tidal torque theory, and that 
the spins of luminous galaxies might be more strongly aligned with the 
surrounding matter than the underlying dark halos. If applied to large scale 
surveys like Sloan Digital Sky Survey (SDSS), our analytic approach will allow 
us to measure accurately the strength of galaxy intrinsic alignment which 
plays a role of statistical error in weak lensing searches and can be used as 
a fossil record to reconstruct cosmology. 

\end{abstract} 

\section{INTRODUCTION}

A flurry of recent cosmological research has been focused on studying the 
galaxy intrinsic alignment.  A question of galaxy intrinsic alignment is 
closely related to the origin of the galaxy angular momentum. 
The standard model for the origin of the galaxy angular momentum is the tidal 
torque theory  \citep{pee69,dor70,whi84}.  According to this theory, the 
galaxy angular momentum is generated by the early tidal interaction with 
the surrounding matter that continued until the moment of the turn-around. 

A generic prediction of the tidal torque theory is the existence of a 
{\it local} correlation between the spin and the matter fields: The spin axes 
of proto-galaxies are preferentially aligned with the intermediate principal 
axes of the tidal shears from the surrounding matter. The degree of this 
galaxy intrinsic alignment must be highest at the moment of the turn-around. 
In the subsequent evolution, the galaxies are likely to lose gradually the 
initial memory about the surrounding matter due to the complicated nonlinear 
effects, with their intrinsic alignments diminishing in strength as a 
consequence. 

An important question in cosmology is whether the intrinsic alignment 
remains at present epoch to a statistically significant degree or not. It is 
important in cosmology since the galaxy intrinsic alignment plays a role 
of systematic error in weak gravitational lensing searches 
\citep{hea-etal00,cro-met00,cat-etal00, cri-etal02,hir-etal04}. 
Besides, it can provide a crucial clue to the unsolved problem of galaxy 
formation \citep{dub92,cat-the96,por-etal02a,por-etal02b,nav-etal04}, and 
be also used as a fossil record to reconstruct cosmology 
\citep{lee-pen00,lee-pen01,lee-pen02}. 

Without knowing the tidal shear field accurately, it is difficult to measure 
the strength of the intrinsic alignment in practice directly from 
observational data.  One possible way is to measure it indirectly from N-body 
simulation data \citep{lee-pen00,hea-etal00,por-etal02a,por-etal02b}. 
However, using N-body data, what one can measure is only the intrinsic 
alignment of underlying dark halos rather than that of the observable luminous 
parts of galaxies.  Although the standard theory of galaxy formation assumes 
that the spin axes of luminous parts align with those of dark halos 
\citep{mo-etal98},  it was hinted by recent gasdynamical simulations 
\citep{che-etal03,nav-etal04} that the luminous parts could conserve the 
initial memory of the intrinsic shear-spin correlation better than the dark 
matter parts, which prevents us from taking the simulation results as real 
ones. \citet{lee-pen02} attempted to measure the strength of intrinsic 
alignment directly from observational data by reconstructing the tidal shear 
field with  the data from the Point Source Catalog of Redshift Survey (PSCz).  
Although they claimed a detection of intrinsic shear-spin correlation, their 
measurement was far from being accurate due to the high statistical noise.

The difficulty in measuring the galaxy intrinsic alignment lies in the fact 
that it is a local correlation between the galaxy spin and the surrounding 
matter fields effective only over a few Mpc in a three dimensional space.  
However, it is  often observed in the universe that the galaxies are 
surrounded by a coherent two-dimensional sheet-like structure which usually 
extends over tens of Mpc.  For such galaxies that are embedded in a sheet, 
one may expect a large-scale coherent orientation \citep{nav-etal04}. 
It does not mean that the existence of cosmic sheets leads logically to the 
expectation of a large-scale coherence in the orientation of the galaxies. 
The two concepts are separate: the galaxy intrinsic alignment is due to 
the tidal torques from the surrounding matter, while sheets form through 
gravitational clustering as first collapsed objects \citep{sha-etal95}. 
Sheets are just one example of the large-scale structures that surround 
and tidally interact with galaxies. The expectation of a coherence in the 
orientation of the galaxies embedded in a sheet comes not from the existence 
of the sheet itself but from the large-scale coherent pattern in the 
two-dimensional matter distribution of the surrounding sheet. Therefore, 
it may be possible to predict the coherent orientation of the galaxy spin axes 
relative the plane of the sheet in the frame of the tidal torque theory. 
Furthermore, it may be also possible to evaluate quantitatively the strength 
of this intrinsic inclination of galaxies by an analytic method.  
The goal of this Letter is to explore these possibilities.

\section{ANALYTIC EVALUATIONS}

\subsection{Key Assumptions}

To evaluate the intrinsic alignment of galaxies embedded in sheets, 
we assume the following: 
\begin{enumerate}
\item 
The galaxy angular momentum, or the galaxy spin vector (${\bf L}$) is 
originated by the tidal shear effect from the surrounding matter at the 
proto-galactic stage, which continued till the moment of the turn-around.   
The growth of the proto-galaxy angular momentum is determined by its  
inertia tensor (${\bf I}$) and the local tidal shear tensor (${\bf T}$). 
The two tensors are strongly but not perfectly correlated with each other 
\citep{lee-pen00,por-etal02a,por-etal02b}.  The slight misalignment between 
the principal axes of the two tensors results in a preferential alignment 
of the proto-galaxy spin axis with the intermediate principal axis of the 
tidal shear tensor (or similarly, the intermediate principal axis of the 
inertia tensor). 
\item
The strength of the galaxy intrinsic alignment with local shears at 
present epoch can be expressed as the following simple quadratic relation 
\citep{lee-pen02}: 
\begin{equation}
\label{eqn:spin}
\langle L_{i}L_{j}\rangle = \frac{1+c}{3}\delta_{ij} - 
c\hat{T}_{ik}\hat{T}_{kj},
\end{equation}
where $\hat{\bf T}$ is the rescaled traceless shear tensor defined as 
$\hat{T}_{ij} \equiv \tilde{T}_{ij}/\vert\tilde{\bf T}\vert$ with 
$ \tilde{T}_{ij} \equiv T_{ij} - {\rm Tr}({\bf T})\delta_{ij}/3$, and $c$ 
is a correlation parameter \citep[see Appendix A in][]{lee-pen02} introduced 
to quantify the strength of the intrinsic shear-spin alignment in the range 
of $[0,1]$.  The value of $c$ cannot be determined 
from 1st physical principles since it includes all the nonlinear effects 
after the moment of the turn-around. One should determine its value 
empirically.  
\item
The probability distribution of galaxy spin vectors, $P({\bf L})$, is 
well approximated as a Gaussian \citep{cat-the96}.
\item
A cosmic sheet is the first collapsed object \citep{sha-etal95}, forming as 
early as the turn-around moment \citep{pau-mel95}.  The minor axis of 
${\bf T}$ at each point on a sheet (i.e., the direction of its maximum 
compression) is almost perpendicular to the plane of the sheet. 
We approximate the lowest eigenvalue of each ${\bf T}$ on a sheet as 
zero coherently.  

\end{enumerate}

\subsection{Distribution of Position Angles}

According to the third hypothesis in $\S 2.1$, the distribution of a galaxy 
spin vector can be written as  
\begin{equation}
\label{eqn:pro}
P({\bf L}) = \frac{1}{[(2\pi)^3 det(M)]^{1/2}}
\exp\left[-\frac{L_{i}(M^{-1})_{ij}L_{j}}{2}\right],
\end{equation}
with the covariance matrix $M_{ij} \equiv \langle L_{i}L_{j}\rangle$ 
is related to ${\bf T}$ by equation (\ref{eqn:spin}). 

Let $\lambda_{1},\lambda_{2},\lambda_{3}$ represent the three eigenvalues 
(in a decreasing order) of the tidal shear tensor ${\bf T}$ at some point 
on the sheet.  Following the fourth hypothesis, we approximate $\lambda_{3}$'s 
identically as zero at all points on the sheet. Therefore, in the frame  
of the principal axis of the tidal shear tensor at each point, 
the polar angle $\theta$ of a galaxy spin vector 
${\bf L} = (L\sin\theta\cos\phi,L\sin\theta\sin\phi,L\cos\theta)$ represents 
the angle of the spin axis relative the plane of the sheet  
(i.e., the galaxy position angle). 
 
To derive the distribution of the galaxy position angles relative to the 
plane of the sheet, we integrate equation (\ref{eqn:pro}) in the principal 
axis frame of the tidal shear as 
$P(\theta) = \int_{0}^{2\pi}\int_{0}^{\infty}P(L,\theta,\phi)L^{2}dL d\phi$, 
and derive the following analytic expression:
\begin{equation}
\label{eqn:ptheta}
P(\theta) = \frac{1}{2\pi}\prod_{i=1}^{3}
\left(1+c-3c\hat{\lambda}^{2}_{i}\right)^{-\frac{1}{2}}\int_{0}^{2\pi}
\left(\frac{\sin^{2}\theta\cos^{2}\phi}{1+c-3c\hat{\lambda}^{2}_{1}} + 
\frac{\sin^{2}\theta\sin^{2}\phi}{1+c-3c\hat{\lambda}^{2}_{2}} + 
\frac{\cos^{2}\theta}{1+c-3c\hat{\lambda}^{2}_{3}}\right)^{-\frac{3}{2}}d\phi
\end{equation}
where $\theta$ is assumed to be in the range of $[0,\pi/2]$. Here the three 
$\hat{\lambda}_{i}$'s ($i=1,2,3$) are the rescaled eigenvalues 
of $\tilde{\bf T}$ constrained by $\sum_{i}\hat{\lambda}^{2}_{i}=1$. 
For the case of $\lambda_{3}=0$, they can be expressed in terms of the two 
eigenvalues $\lambda_{1}$ and $\lambda_{2}$ as 
\begin{equation}
\label{eqn:hatl}
\hat{\lambda}_{1} = \frac{2\lambda_{1}-\lambda_{2}}
{\sqrt{6(\lambda^{2}_{1}-\lambda_{1}\lambda_{2}+\lambda^{2}_{2}})}, \quad
\hat{\lambda}_{2} = \frac{2\lambda_{2}-\lambda_{1}}
{\sqrt{6(\lambda^{2}_{1}-\lambda_{1}\lambda_{2}+\lambda^{2}_{2}})}, \quad
\hat{\lambda}_{3} = \frac{-(\lambda_{1}+\lambda_{2})}
{\sqrt{6(\lambda^{2}_{1}-\lambda_{1}\lambda_{2}+\lambda^{2}_{2}})}
\end{equation}

Equations (\ref{eqn:ptheta})-(\ref{eqn:hatl}) show how the probability 
distribution of the galaxy position angles, $P(\theta)$,  depend on the 
eigenvalues of $\lambda_{1}$ and $\lambda_{2}$ as well as the correlation 
parameter $c$.  The conditional distributions of the two eigenvalues 
provided that $\lambda_{3}=0$ can be derived by using the Bayes Theorem: 
$p(\lambda_{i}|\lambda_{3}=0) = 
[p(\lambda_{i},\lambda_{3}=0)/p(\lambda_{3}=0)]$ ($i=1,2$). Here  
the two-point joint distribution $p(\lambda_{i},\lambda_{3})$ and 
the one-point distribution $p(\lambda_{3})$ are all obtained 
analytically by \citet{lee-sha98}.  Using these results, we derive 
\begin{eqnarray}
\label{eqn:l1}
p(\lambda_{1}|\lambda_{3}=0)&=&\frac{27(3\sqrt{3}+\sqrt{2})}
{16\sqrt{\pi}\sigma}\bigg{\{}\frac{3\lambda^{2}_{1}}{\sigma^{2}}
\exp\left(-\frac{3\lambda^{2}_{1}}{\sigma^{2}}\right) + 
\frac{\lambda^{2}_{1}}{\sigma^{2}}
\exp\left(-\frac{9\lambda^{2}_{1}}{2\sigma^{2}}\right) + \nonumber \\
&&\frac{\sqrt{3\pi}\lambda_{1}}{12\sigma}
\left(\frac{9\lambda^{2}_{1}}{\sigma^{2}}-8\right)
\exp\left(-\frac{45\lambda^{2}_{1}}{16\sigma^2}\right)
\left[{\rm erf}\left(\frac{3\sqrt{3}\lambda_1}{4\sigma}\right) + 
{\rm erf}\left(\frac{\sqrt{3}\lambda_1}{4\sigma}\right)\right]\bigg{\}}, \\
\label{eqn:l2}
p(\lambda_{2}|\lambda_{3}=0)&=&\frac{27(3\sqrt{3}+\sqrt{2})}
{16\sqrt{\pi}\sigma}\bigg\{\frac{\lambda^{2}_{2}}{\sigma^{2}}
\exp\left(-\frac{9\lambda^{2}_{2}}{2\sigma^{2}}\right) + \nonumber \\ 
&&\frac{\sqrt{3\pi}\lambda_{2}}{12\sigma}\left(8-\frac{9\lambda^{2}_{2}}
{\sigma^{2}}\right)
\exp\left(-\frac{45\lambda^{2}_{2}}{16\sigma^2}\right){\rm erfc}
\left(\frac{3\sqrt{3}\lambda_2}{4\sigma}\right)\bigg\}, 
\end{eqnarray}
where $\sigma$ is the rms fluctuation of the linear density field. 

\begin{figure}
\plotone{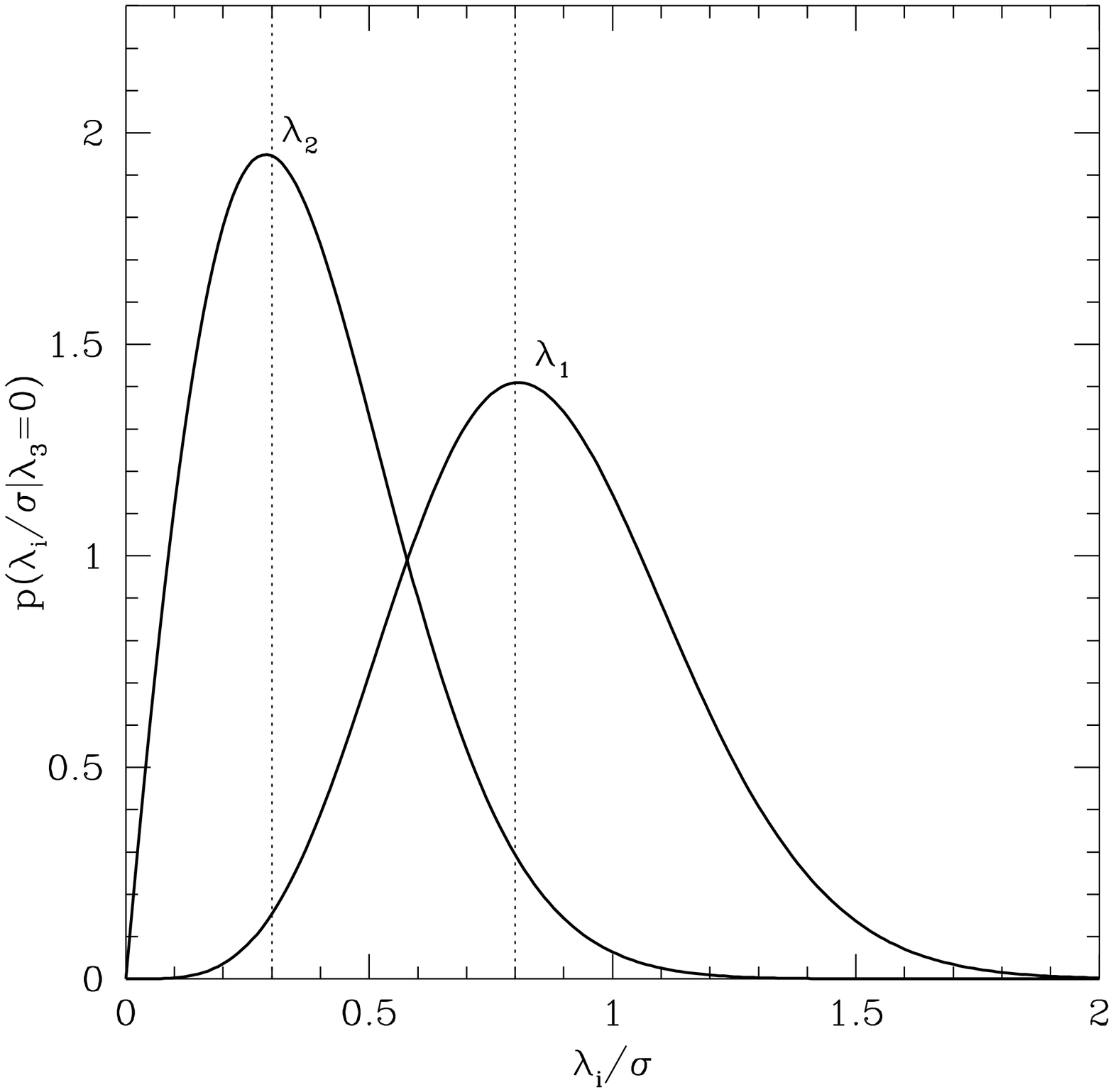}
\caption{Conditional probability density distributions of the largest 
and the second largest eigenvalues of an intrinsic shear tensor provided 
that its smallest eigenvalue has the value of zero. The dotted line 
locates the most probable eigenvalue for each case. 
\label{fig:cond}}
\end{figure}

Figure \ref{fig:cond} plots the above conditional distributions 
(eqs.[\ref{eqn:l1}]-[\ref{eqn:l2}]). The dotted lines locate the most probable 
values of the two eigenvalues provided that $\lambda_{3}=0$: 
$\lambda_{1}^{\rm max} \approx 0.8\sigma$; 
$\lambda_{2}^{\rm max} \approx 0.3\sigma$. 
We also determined the conditional mean and the standard deviation of each 
eigenvalue from equations (\ref{eqn:l1})-(\ref{eqn:l2}):  
$\bar{\lambda}_{1}=0.86\sigma$, $\sigma_{\lambda_{1}}=0.28\sigma$; 
$\bar{\lambda}_{2}=0.38\sigma$, $\sigma_{\lambda_{2}}=0.21\sigma$.  

\begin{figure}
\plotone{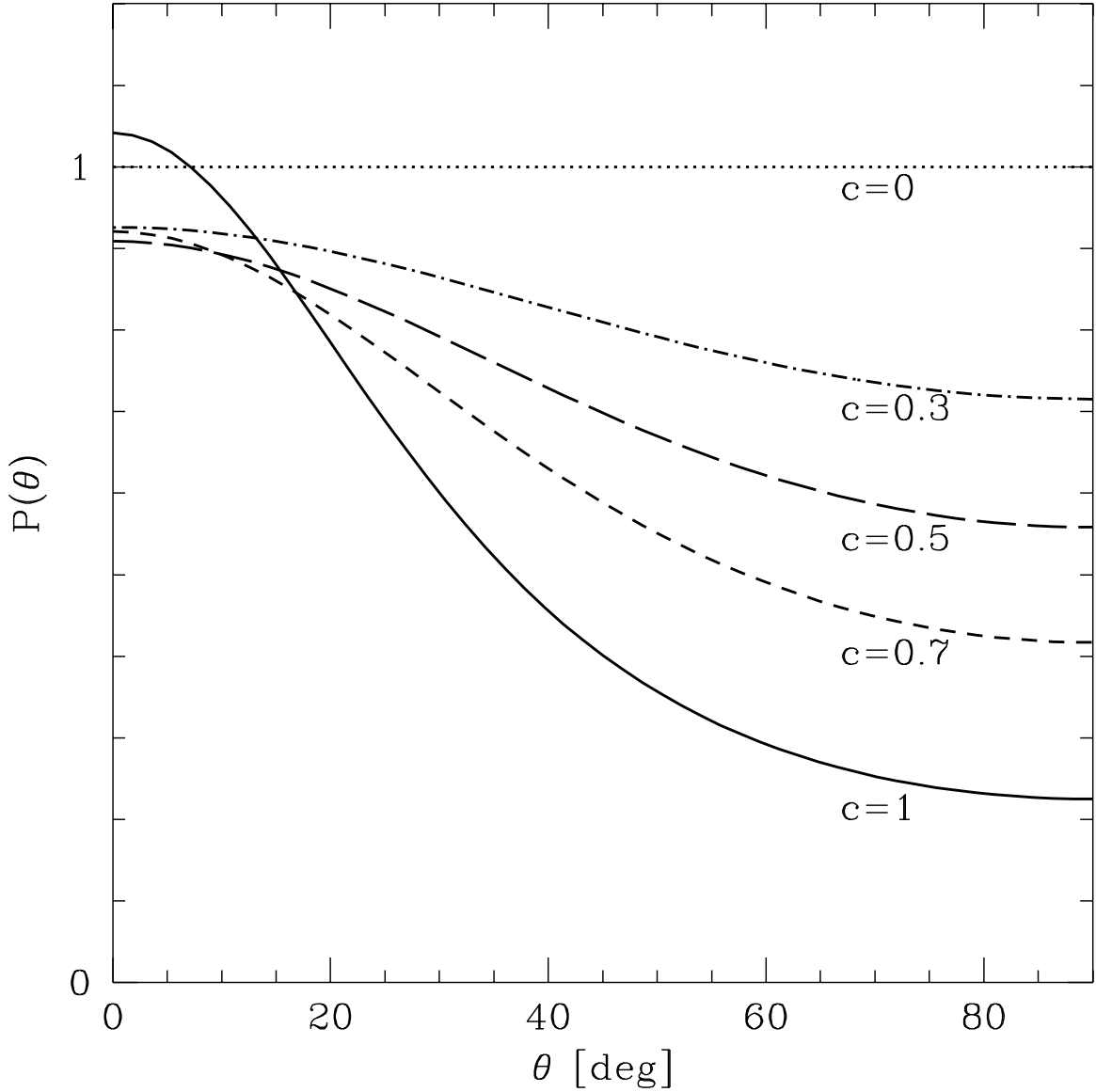}
\caption{Probability density distributions of the angles of the minor axes 
of galaxies embedded in a two-dimensional sheet relative the plane of the 
sheet. The solid, dahsed, long-dashed, the dot-dashed, and dotted lines 
represent the cases of the correlation parameter $c = 1,0.7,0.5,0.3$ and 
$0$, respectively,
\label{fig:proc}}
\end{figure}

Putting $\lambda_{1}^{\rm max}$ and $\lambda_{2}^{\rm max}$ into 
equation (\ref{eqn:ptheta}), we can investigate how the distribution of 
the galaxy position angles relative to the plane of the sheets varies with 
the value of the correlation parameter $c$. Figure \ref{fig:proc} plots 
$P(\theta)$ for five different cases of $c=1,0.7,0.5,0.3$ and $0$ 
(solid, dashed, long-dashed, dot-dashed and dotted lines, respectively). 
As expected, for the case of $c=0$, $P(\theta)$ is uniform, and the larger 
the value of $c$ is, the more sharply does $P(\theta)$ increase in the 
small-angle section. 
In other words, the stronger the intrinsic shear-spin correlation is, 
the more inclined are the galaxy spins relative to the sheet. 
Equation (\ref{eqn:ptheta}) also allows us to evaluate the average position 
angles: $\bar{\theta} = \int_{0}^{\pi/2}\theta P(\theta)d\theta$. We find 
$\bar{\theta} = 29$ in unit of degree, if $c = 0.9$ (see in $\S 3$).  

\section{THEORY VS. OBSERVATION}

A vivid example of such galaxies that are embedded in a two-dimensional 
sheet comes from the neighborhood of our Milky Way where nearby spirals 
are assembled in the Local Super Cluster \citep{RC3}. Although there were 
some reports of detecting a large-scale coherent orientation of the observed 
nearby spirals relative to the Local Super Cluster 
\citep[e.g.,][]{gre-etal81,hel-sal82,fli-god86,gar-etal93}, 
these previous reports often suffer from small sample sizes, hardly being 
established as compelling evidences. 

However, very recently, \citet{nav-etal04} used a relatively large sample 
of nearby galaxies from the Principal Galaxy Catalog 
(PGC, Paturel et al. 1997), and estimated the number distribution of nearby 
disk galaxies as a function of the supergalatic position angles. 
They found that the spin axes of nearby edge-on spirals ($cz < 1200$ km/s) 
are strongly inclined relative to the supergalactic plane, and concluded 
that the observed result is consistent qualitatively with the tidal torque 
theory.  Yet, they did not make any quantitative comparison of the observed 
result with the predictions of the tidal torque theory. 
 
\begin{figure}
\plotone{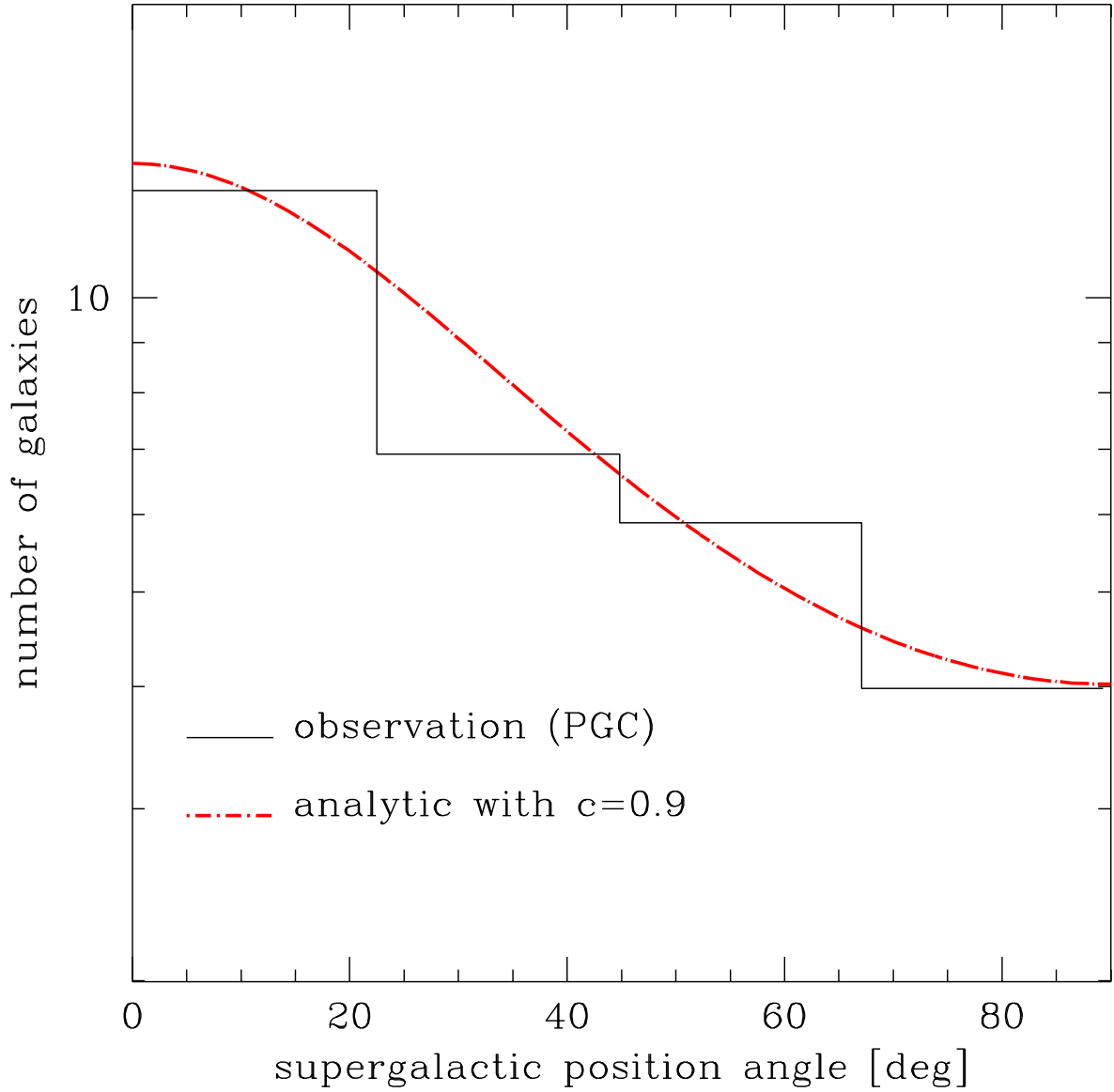}
\caption{Number distributions of supergalactic position angles of 
spirals embedded in a sheet. The dot-dahsed curve represents the 
theoretical prediction assuming $c=0.9$, while the histogram is  
the observational result from PGC catalog (see Fig. 2 in \citet{nav-etal04}
\label{fig:obs}}
\end{figure}

To make a quantitative comparison, we evaluate the number distribution 
of galaxies embedded in sheets with equation (\ref{eqn:ptheta}). 
Figure \ref{fig:obs} plots the results. The (red) dot-dashed line represents 
our theoretical prediction with the choice of $c=0.9$ (the best-fit value), 
while the histogram corresponds to the observational result from PGC 
\citep[see Fig. 2 in][]{nav-etal04}.  Obviously, the theoretical curve agrees 
with the histogram very well. \citet{nav-etal04} determined the average value 
of the position angles from the observed edge-on disk galaxies to be 
$\bar{\theta}_{obs} \approx 25$ in unit of degree, which is in agreement of 
the theoretical value that we obtained assuming $c=0.9$ in $\S 2$. 

It is worth noting that the best-fit value of $c=0.9$ is three times higher 
than that found in N-body simulation \citep{lee-pen00,lee-pen02}.
It may imply that the luminous parts of galaxies tend to keep the initial 
memory of the surrounding matter field better than their dark matter 
counter parts, consistent with the results from recent gasdynamical 
simulations \citep{che-etal03,nav-etal04}. 

\section{CONCLUSION}

We have predicted analytically a large-scale coherence in the orientation 
of galaxies embedded in sheets using the tidal torque theory. Our analytic 
model reproduces the observed inclinations of nearby spiral galaxies in 
the Local Super Cluster remarkably well, providing a physical quantitative 
understanding of the observables.  It should be, however, emphasized here 
that the prediction of the coherent orientation of galaxies embedded in a 
sheet should not be taken as identical to the existence of the sheet itself. 
The coherent orientation of the galaxies is caused by the tidal torquing from 
the surrounding matter, which is different from the formation of cosmic 
sheets in the universe. 

By comparing our analytic prediction with the observational result, we 
also measured quantitatively the strength of the galaxy intrinsic alignment, 
and found that the luminous galaxies seem to conserve their angular 
momentum better than the underlying dark halos after the moment of the 
turn-around.   Our final conclusion is this: our analytic calculation, 
if applied to observational data from large scale surveys like Sloan Digital 
Sky Survey (SDSS), will allow us to measure accurately the strength of galaxy 
intrinsic alignment that will have an impact in weak gravitational searches, 
and can also be used as a fossil record to reconstruct cosmology. 

\acknowledgments
We thank C. Park and U. Pen for useful discussions. We also thank the 
referee for helpful suggestions. This work is supported by the research 
grant of the Korea Institute for Advanced Study (KIAS).

\clearpage

\clearpage

\clearpage

\end{document}